# PROJECT ICARUS: PRELIMINARY THOUGHTS ON THE SELECTION OF PROBES AND INSTRUMENTS FOR AN ICARUS-STYLE INTERSTELLAR MISSION

**I. A. Crawford**

*Department of Earth and Planetary Sciences, Birkbeck College London, Malet Street, London, WC1E 7HX, UK*

**Abstract**

In this paper we outline the range of probes and scientific instruments that will be required in order for Icarus to fulfill its scientific mission of exploring a nearby star, its attendant planetary system, and the intervening interstellar medium. Based on this preliminary analysis, we estimate that the minimum total Icarus scientific payload mass (i.e. the mass of probes and instruments which must be decelerated to rest in the target system to enable a meaningful programme of scientific investigation) will be in the region of 100 tonnes. Of this, approximately 10 tonnes would be allocated for cruise-phase science instruments, and about 35 tonnes (i.e. the average of estimated lower and upper limits of 28 and 41 tonnes) would be contributed by the intra-system science payload itself (i.e. the dry mass of the stellar and planetary probes and their instruments). The remaining ~55 tonnes is allocated for the sub-probe intra-system propulsion requirements (crudely estimated from current Solar System missions; detailed modelling of sub-probe propulsion systems will be needed to refine this figure). The overall mass contributed by the science payload to the total that must be decelerated from the interstellar cruise velocity will be considerably more than 100 tonnes, however, as allowance must be made for the payload structural and infrastructural elements required to support, deploy, and communicate with the science probes and instruments. Based on the earlier Daedalus study, we estimate another factor of two to allow for these components. Pending the outcome of more detailed studies, it therefore appears that an overall science-related payload mass of ~200 tonnes will be required. This paper is a submission of the Project Icarus Study Group.

**Keywords:** Interstellar travel; exoplanets; interstellar medium; stellar properties; planetary science; astrobiology

## 1. Introduction

The Icarus study [1,2] is tasked with designing an interstellar space vehicle capable of making in situ scientific investigations of a nearby star and its accompanying planetary system. This paper presents an initial consideration the range of probes and scientific instruments that will be required in order for Icarus to fulfill its scientific mission. It is important to realize that at this stage in the project any such analysis can only be very preliminary, as decisions regarding the actual complement of scientific instruments carried by an interstellar mission will depend on the following:

(i) The available mass, power, and communications bandwidth budgets; and

(ii) The architecture of the target star and planetary system (especially the number and type of planets present, including any observations of possible biosignatures that may have been made by solar system-based astronomical observations).

For the purposes of the Icarus project, it is expected that understanding of (i) will mature as the design progresses, whereas (ii) will be satisfied by the definition of a hypothetical, but plausible, planetary system at a later stage of the project.

Although neither of these key prerequisites are currently defined, it is nevertheless considered worthwhile at this stage of the project to identify at least the kinds of probes and instruments, and their approximate numbers, that would be required by an Icarus-type interstellar mission conducting an initial scientific investigation of a nearby exoplanetary system.

## 2. Top-level scientific objectives

As discussed by Crawford [3], and building on earlier work by Webb [4] in the context of the Daedalus study, the Icarus science objectives include:

(1) Science to be conducted on route. This includes studies of the interstellar medium (ISM) encountered traversed on the way to the target star, and also physical and astrophysical studies which can make use of the Icarus vehicle as an observing platform;

(2) Astrophysical studies of the target star itself, or stars, if a multiple system is selected;

(3) Planetary science studies of any planets in the target system, including moons and large asteroids of interest; and

(4) Astrobiological/exobiological studies of any habitable (or inhabited) planets which may be found in the target planetary system.

These broad science areas may themselves be sub-divided into a number of different areas of investigation. These are summarised in Table 1.

The remaining sections of this paper outline the kinds, and approximate numbers, of sub-probes and accompanying instruments that would be required to address these different scientific questions.

## 3. Science probes and instruments

3.1 Science conducted *en route*

The main scientific investigations to be conducted *en route* will be studies of the interstellar medium (ISM) between the Sun and the target star system. The sizes, physical properties, and locations of the low-density cloudlets known to be present in the vicinity of the Sun, and the nature of the lower density inter-cloudlet medium, are of particular interest [5]. Key instruments for these studies would be:

- A dust analyser (to determine both dust masses and compositions)

- Gas analyser(s) (to measure densities, ionisation states, and composition of the gas phase interstellar medium – probably a suite of several instruments in practice)

- Magnetometer(s) and plasma wave instruments

- High-energy (cosmic ray) detectors

The instruments required to study the structure of the interstellar medium should have the highest priority for cruise-phase science. These instruments would be similar to those routinely flown on outer Solar System missions (e.g. Cassini; [6]) and proposed for interstellar precursor missions (e.g. [7]). They could be mounted on the main vehicle, but booms (or free-flying sub-probes) may be desirable to insulate them as much as possible from electromagnetic and other interference from the vehicle. Thought will also have to be given to protection from interstellar dust impacts.

If Icarus is also to be used as a platform for detecting 'exotic' (e.g. dark matter) particles then additional instrumentation may be required. The Alpha Magnetic Spectrometer [8], currently mounted on the ISS, could provide an example, although in practice an innovative new instrument would probably need to be designed. If Icarus was to be used as a platform for optical or radio parallax studies, appropriate instrumentation (i.e. optical or radio telescopes) would need to be defined and a allowance made in the appropriate mass budget.

Table 1 also identifies studies of the Solar System's heliosphere, Kuiper Belt and Oort Cloud as possible targets for investigation. These three areas of investigation are rather different. The instrumentation described above for the study of the ISM will also be sufficient for heliospheric studies, and indeed the boundary between the heliosphere and the local ISM is of compelling scientific interest [5]. However, it is not considered here that Kuiper belt objects (KBO) should be a high priority for Icarus, partly because pre-Icarus precursor missions are likely to have made in situ observations of these objects, and partly because refining the Icarus trajectory to pass close a KBO may overly complicate the flight profile. The Oort cloud is so thinly populated that it is unlikely that Icarus will pass close enough to an Oort cloud comet to make in situ observations, and in any case the vehicle will probably be travelling at or close to its cruise velocity, making such observations difficult (and potentially dangerous).

3.2 Studies of target star(s)

Studies of the target star(s) will be of compelling astrophysical interest. However, it must be noted that for these very nearby stars a lot can, and will, be learned from astronomical observations made from Earth. It follows that *in situ* studies should be designed to complement these astronomical studies, and be used to obtain information that cannot plausibly be obtained using telescopic observations from the Solar System. Examples include *in situ* measurements of stellar wind and magnetic field, high-spatial resolution observations of stellar photospheres, active regions and coronae, and ultra-high precision measurements of stellar mass, radii and luminosity.

The choice of suitable instruments should be guided by instruments on existing solar missions such as the Solar and Heliospheric Observatory (SOHO, [9]), Solar Terrestrial Relations Observatory (STEREO, [10]), and the forthcoming Solar Orbiter [11]. Standard instruments would include:

- Magnetometers

- Charged particle (stellar wind) detectors

- X-Ray, UV and visible imaging systems

- Instrument to measure total stellar luminosity to high precision (essentially the stellar 'solar constant') and any variability therein.

Most of these instruments could be mounted on the main spacecraft bus rather than on sub-probes, and some of them (excluding the imaging instruments) could probably be the same as those used during the cruise phase to study the ISM. That said, the main vehicle is unlikely to be operating at an optimal distance from the star for many stellar physics studies, and in any case the STEREO mission [10] has demonstrated the value of making simultaneous observations of solar (here stellar) activity from different angles. It follows that at least one, and ideally two, sub-probe/s dedicated to stellar observations in addition to instruments on the main spacecraft would be desirable (and at least two, and ideally four, such stellar-physics sub-probes if a binary star such as alpha Centauri is the target).

3.3 Planetary science

The requirements for planetary probes will depend on the number and types of planets present in the target system. It is anticipated that, at least for planets of Moon-size and larger, by the time Icarus is launched this information will have been provided by Solar System-based astronomical observations. In what follows, we identify some generic instrumentation for different modes of planetary science investigations (although closely related, specifically astrobiology-related investigations are discussed in Section 3.4).

3.3.1 Planetary orbiters

Planetary orbiters provide an efficient means of mapping planets, determining surface and atmospheric composition, and making top-level inferences about their geological evolution. In general, planetary mapping orbiters should be placed in polar orbits to maximize surface coverage, and this will either require each planet's polar axis to be identified before orbital insertion, or each orbiter to carry sufficient fuel to effect orbital plane changes. The instrumentation required on planetary orbiters will depend somewhat on the nature of the planet (e.g. gas giant or rocky planet), but orbiters designed for giant planets should also be able to explore orbiting moons (as for the Galileo and Cassini missions). Instrumentation will also depend on the presence or absence of an atmosphere, and the extent to which a planet may be totally cloud covered (like Venus and Titan). These may be difficult to determine in advance, so a flexible instrument suite would be desirable.

Typical instruments for planetary orbiters would include:

- High-resolution optical imaging system

- High-resolution imaging UV-VIS-IR mapping spectrometer(s)

- X-Ray fluorescence (XRF) or gamma-ray spectrometer (airless bodies only)

- Laser altimeter (airless bodies or planets with transparent atmospheres)

- Synthetic aperture radar (primarily for mapping cloud covered planets)

- Magnetometer

3.3.2 Atmospheric probes: giant planets

Here the principal interest will be in making in situ measurements of the chemical composition of the atmosphere. A good model would be the Galileo Jupiter atmospheric entry probe [12], although if long-term monitoring were considered desirable balloons or aircraft (drones) might be considered. Principle instruments for such probes would include:

(1) Mass spectrometer(s)

(2) Nephelometer

(3) Thermometers, barometers, etc

Line-of-sight communication with an orbiter, or with the main vehicle, will be required for data transfer. Doppler tracking of the signal will permit measurements of wind speed at different depths.

3.3.3 Atmospheric probes: rocky/icy planets

To a first approximation the kinds of instruments required to probe the atmospheres of solid planets are the same as required for giant planets (see above), but in this case balloons and/or aircraft would appear to be especially desirable. This is because, in addition to measuring atmospheric parameters, they could assist in surface exploration (e.g. by obtaining high-resolution images and spectral data of the surface, something which would be especially important if the surface was obscured from orbital investigation by clouds or hazes). In this case, in addition to the instruments designed to characterize the atmosphere (listed in Section 3.3.2 above) additional instruments for surface observations could include:

- High-resolution multi-spectral imaging system

- UV-VIS-IR mapping spectrometer(s)

3.3.4 Rocky/icy planet surfaces: hard landers/penetrators

From a planetary science point of view, much of the information we would like to obtain about solid planets will require *in situ* measurements made by contact instruments at the surface. Obtaining some of this information will require sophisticated instruments to be soft-landed on the planetary surface, and many will require rover-facilitated mobility (as discussed below). However, a lot of valuable top-level geophysical and geochemical information can be obtained by suitably instrumented hard landers or penetrators, which would be dropped from orbit and land intact on, or embed themselves a few metres below, the surface. Penetrators are likely to be especially efficient at emplacing network geophysical instruments (e.g. seismometers and heat-flow probes), as several penetrators could be targeted at each planet of interest. This mode of deploying instruments will be most effective on airless bodies, but might be adapted for planets with atmospheres as well.

Examples of suitable penetrator studies include the MoonLITE [13] and LunarEX [14] concepts proposed for lunar exploration, and a similar concept proposed for Europa [15]. As demonstrated in these studies, examples of the kind of scientific instruments which could be efficiently deployed using penetrators include:

- Seismometers

- Heat-flow-probes

- In situ geochemical sensors (e.g. mass spectrometers and X-ray fluorescence spectrometers)

Note that at least intermittent line-of-sight communication with an orbiting satellite will be required for data downlink from penetrators.

3.3.5 Rocky/icy planet surfaces: soft landers/rovers

Some important planetary science investigations, especially those relating to geology and astrobiology, will require larger and more complex instrumentation than can plausibly be emplaced by penetrators. Moreover, many of these investigations would benefit from mobility, which implies the ability to land rovers on planetary surfaces. A good state-of-the-art rover, that is equipped with appropriate instrumentation, is NASA's Mars Science Laboratory [16].

Examples of the kinds of instruments required for surface geological and environmental investigations include:

- Multi-spectral imaging system

- Weather station (e.g. temperature, humidity, wind speed, etc)

- Rock abrasion tool and/or rock splitter

- High-resolution close-up ('hand lens') imager

- *In situ* geochemical instruments (e.g. XRF, Raman, and/or laser-induced breakdown spectrometers)

- Mass spectrometer(s). Note that different mass spectrometers may be required for geochemical and astrobiological investigations, as the former will mostly be concerned with trace element concentrations in rock samples, whereas the latter will be concerned with identifying complex organic molecules; we return to the latter in Section 3.4.

- *In situ* rock dating capability (using some combination of the above instruments, as demonstrated recently by Farley et al. [17])

- Drill (for sub-surface sampling; depth TBD, but ideally several metres)

- Geophysical package (e.g. seismometer, heat-flow probe).

As was the case for penetrators, at least intermittent line-of-sight communication with an orbiting satellite will be required for data downlink from surface landers.

3.3.6 Small bodies: asteroids and comets

By analogy with the Solar System, we would expect that the minor bodies (i.e. comets and asteroids) of exoplanetary systems will contain valuable information relating to the age and formation history of the system. This will make such objects valuable targets of investigation. Examples of Solar System missions to such bodies include the Dawn mission to Vesta and Ceres [18] and the Rosetta mission (with Philae lander) to Comet 67P/Churyumov–Gerasimenko [19]. Instruments would be a sub-set of those identified above for the investigation of planetary surfaces.

Note that obtaining the *age* of primitive cometary and/or asteroidal material through radiometric dating will be especially valuable from an astrophysical perspective. Such measurements would provide an independent estimate of the age of the parent star, and therefore help calibrate age estimates based on stellar evolution models for a star (or stars) having a spectral type other than the Sun. The recent results of Farley et al. [17] give some confidence that, by the time an Icarus-style interstellar mission can be mounted, such *in situ radiometric* dating will be possible. Note, however, that it will likely require a more sophisticated and more massive cometary/asteroid lander than the ~100 kg Philae probe.

3.4 Astrobiology

Requirements for astrobiological investigations are closely related to those of planetary science. In particular, they will likely require the soft-landing of rover-facilitated mobile instruments as discussed in Section 3.3.5). Presumably dedicated astrobiology instruments would only be targeted those planets identified in advance as being good candidates for habitability (e.g. possessing liquid water and a life-friendly atmospheric composition), or for which atmospheric biomarkers have already been detected spectroscopically (either from Earth or from previously deployed orbiting spacecraft). In the event that biosignatures have been detected from Earth prior to launch it is likely that this would dominate the entire Icarus scientific investigation of the target system, and that the scientific payload would be tailored to its further investigation (possibly at the expense of some of the other scientific objectives outlined above). Even in the absence of the prior detection of actual biosignatures, it is likely that astronomical observations from the Solar System will have identified potentially habitable planets in the target system, if such exist, and that this information will also inform the particular choice of astrobiology experiments to be included in the payload.

In the absence of such prior information, we here outline generic astrobiology investigations that might be suitable for the investigation of potentially habitable planets in the target system. Good models for astrobiological instrument suites include the Viking biology package [20] (although it would be possible to design a more sophisticated package today, and ideally more tailor-made to the specific target environment to the extent that this can be known in advance), the Pheonix Lander high-resolution microscope [21], and the Surface Analysis at Mars (SAM, [22]) and Urey [23] instruments designed for Mars Science Laboratory and ExoMars, respectively.

Appropriate instruments would the first eight bullet points identified in Section 3.3.5 above as being required for planetary science investigations, but with the addition of the following:

- High-resolution microscope

- Mass spectrometer(s) for detection and characterization of complex organic molecules and carbon (and perhaps other element) isotope ratios

- Measurements of temperature, pH and redox state (Oxidation-Reduction Potential, ORP) of liquid water (or any other liquids) found in the vicinity of the landing site

- 'Wet' biology experiments to identify and characterize active metabolism and metabolic products (e.g. loosely based on the Viking biology package, but with updated (and ideally specially tailored) experimental protocols)

- Some kind of metagenomic analysis might be desirable, but probably impossible without precise knowledge of indigenous genetic processes.

If the target planet is wholly or partially covered in liquid water then, given the importance of liquid water to biology as we understand it, it will be desirable to deploy some of these instruments on a mobile floating platform (i.e. a boat) and to obtain and analyse water samples. In such a situation, a water (liquid) sampler would replace the drill specified in Section 3.3.5.

**4. Number or probes required and implications for payload mass**

The number of probes required to conduct a thorough investigation of the target planetary system will depend on number and types of planets present. In practice, this is likely to be determined before the launch of an Icarus-style interstellar mission by astronomical observations made from the Solar System. Such observations will also inform the relative balance between giant planet and rocky planet investigations, and the weighting given to astrobiological investigations.

In the absence of such knowledge it is only possible to make some very rough and preliminary estimates of the minimum number of probes that will be required as follows:

- Stellar orbiters: 2-4

- Planetary orbiters: 6-8 (say one per planet; may need more)

- Giant planet entry probes/balloons: 2-4 (at least one per giant planet)

- Rocky planet atmospheric vehicles/balloons: 6-8 (ideally at least two for each rocky planet with an atmosphere)

- Planetary soft landers: 6-8 (ideally at least two for each rocky planet; at least half should be equipped for astrobiological investigations and at least two designed to function on water)

- Probes to explore/date minor bodies such as asteroids and/or comets: 2-4

- Low mass penetrators for simple in situ geochemical and geophysical studies of airless planets, moons and asteroids: 30-50

It is instructive to estimate the total mass of the above scientific payload. Taking SOHO (dry mass 1.35 tonnes [24]) as a model for a stellar orbiter; Cassini (mass 2.15 tonnes, excluding the Huygens probe [6]) as a high-performance planetary orbiter; the Galileo Entry Probe (mass 0.34 tonnes [12]) as a typical planetary atmospheric probe); Mars Science Laboratory (mass 0.90 tonnes [16]) as a high performance planetary rover with astrobiology capability); Rosetta (mass 1.33 tonnes including Philae lander [19]) as an example asteroid/comet probe; and the LunarEX penetrators (mass 0.04 tonnes, including descent modules [14]) as typical of this type of delivery system, we arrive at a total probe mass in the range of 29 to 43 tonnes for the lower and upper limits for the number of probes given above. For ease of reference, this information is summarized in Table 2.

Note however, that these estimates, although they include the masses of the scientific instruments, communications systems, electronics and power systems, and supporting structure integrated into these various spacecraft, they do not include any fuel or propulsion systems for intra-system maneuvering, orbit insertion, or, where required, landing. These additional masses will be significant. For example, Cassini was launched with 3.1 tonnes of propellant for orbital insertion and maneuvering within the Saturn system [6], which significantly exceeds the mass of the spacecraft itself and brings the total Cassini mass to 5.2 tonnes (i.e. about 2.5 times the dry mass of the spacecraft). Similarly, although the mass of the MSL rover is 'only' 0.9 tonnes, the total launch mass (which of course had to include the aeroshell, heatshield, parachutes and skycrane used for entry, descent and landing) was 3.9 tonnes [16], or 4.3 times the mass of the rover alone.

Clearly, the masses that must be allowed for intra-system maneuvering, and landing where appropriate, will depend on the technologies and energy sources adopted, and this should be a priority for further investigation. Certainly it will be important to adopt an exploration strategy which minimizes the energy requirements for intra-system maneuvering, for example by adopting an approach whereby probes are delivered to their targets from the main vehicle on Hofmann transfer orbits [25]. In the absence of any better information, we here follow the Cassini example and multiply the dry probe masses given above by a factor of 2.5 to allow for whatever propulsion system is needed to transport them to where they need to operate (in the full knowledge that this may be conservative for some probes, e.g. stellar orbiters, and optimistic for others, e.g. planetary landers). This brings our estimate of a minimum stellar and planetary probe mass to 70 to 103 tonnes.

We still have to add the mass of instruments required for cruise phase science discussed in Section 3.1. The basic measurements of fields and particles are similar to those already made by outer solar system spacecraft such as Cassini, so allowing an additional 2.0 tonnes (i.e. approximately the total mass of the Cassini Orbiter) for these instruments would appear to be conservative. However, if it is desired to make use of the interstellar vehicle as a platform for more exotic studies during the cruise phase (e.g. dark matter searches or astronomical observations) additional mass must be allowed for these instruments. As a baseline we here add another 8 tonnes for these unspecified instruments (this may also be conservative, but note that the Alpha Magnetic Spectrometer on the ISS, which might act as a proxy for an advanced exotic particle detector, has a mass of 8.5 tonnes [8]).

This brings the estimated total mass of probes and scientific instruments to lie within the range of 80 to 113 tonnes. Given all the uncertainties, it seems safe to conclude that an Icarus-style interstellar mission could perform a scientifically valuable exploration of a nearby exoplanetary system if it is able to deliver a total scientific payload of order 100 tonnes to the target system.

Of course this scientific payload is only a sub-set of the mass of the total Icarus payload which must be decelerated into the target star system from the vehicles interstellar cruise velocity. The latter will include many additional elements, including (but not limited to): payload-committed structure and power supplies; the main computer and data management systems; intra-system and system-Earth communications systems; the payload dust protection system (or at least that part of it designed to protect the payload as it enters the target system); and whatever fuel and thrusters may be required to maneuver the main vehicle itself within the target system. In his analysis of the Daedalus payload Webb [4; see his Table 1] found the total payload mass to be roughly twice the mass allocated for scientific probes (although he excluded the main vehicle's power supplies and dust protection system in this estimate).

Using this as a baseline, we arrive at a total Icarus payload mass of the order of 200 tonnes for a reasonably complete scientific exploration of a nearby star, its planetary system, and the intervening interstellar medium. Needless-to-say, a larger payload mass would significantly enhance the scientific return

## 5. Conclusions and recommendations

We have presented some initial considerations on the selection of probes and instruments that would be required to make a scientifically useful exploration of a planetary system orbiting a nearby star. In practice, the scientific payload will be tailored to both the capabilities of the interstellar vehicle and the architecture of the particular planetary system to be explored. Information on the latter is likely to be provided by astronomical observations from the Solar System long before an Icarus-style interstellar vehicle is constructed. Nevertheless, based on how we have explored our own Solar System, it is possible to consider a generic scientific payload able to address the top-level science requirements for such an expedition. This has led to the following conclusions:

(1) Based on existing spacecraft and instruments, the dry mass of probes required to perform a minimally useful exploration of a target system containing 1-2 stars and 6-8 planets (such as may be appropriate for number of nearby star systems, including alpha Centauri A/B [25,26]) will probably be in the range of 28-41 tonnes (mid-range: 35 tonnes; Table 2). While this might be reduced by improved technology (especially the likely miniaturization of instruments over the coming century), it seems best to use this figure in order to be conservative (and any mass savings made as a result of improved technology would probably be better used to increase the range of science that could be performed rather than reducing the scientific payload mass).

(2) Allowance must be made for transporting the scientific probes to the locations in orbit about, or on the surfaces of, planets to be investigated. This will require propulsion systems and fuel requirements not considered in the above mass estimate. It is not possible to estimate the mass that must be allocated for probe deployment and intra-system maneuvering until the probe propulsion systems are defined. This should be the basis of a future study. Here, based on chemically propelled Solar System probes, we estimate (perhaps conservatively) that this will increase the total mass required for science probes by a factor of 2.5, bringing it into the range 70-103 tonnes (mid-range: 87 tonnes). Adding an estimated 10 tonnes for cruise phase science, brings the total to 80-113 tonnes (mid-range: 97 tonnes).

(3) Given all the uncertainties, we conclude that an Icarus-style interstellar mission could perform a scientifically valuable exploration of a nearby exoplanetary system if it is able to deliver a total scientific payload of order 100 tonnes to the target system.

(4) In addition to the science payload itself, allowance must be made for the structural and infrastructural elements required to support, deploy, and communicate with the science probes and instruments. Based on the earlier Daedalus study [4], we estimate this to be an additional factor of two, resulting in a total minimum science-related payload mass (i.e. the total science-related payload which must be decelerated to rest in the target system) in the region of 200 tonnes. However, we stress that this must be verified by a detailed integrated study of all the payload systems.

(5) Demonstrating that a mainly fusion-based propulsion system is capable of delivering a science-related payload mass of order 200 tonnes to rest at a nearby star, with a total travel time of under 100 years (as specified in the Icarus Terms of Reference [27]), is therefore an important objective for on-going studies of the Icarus propulsion system.

(6) As 200 tonnes appears to be close to the minimum science-related payload mass required to meet the science objectives, and as de-scoping the science requirements will surely compromise the justification for investing in something as complex and expensive as an interstellar spacecraft, if the Icarus study indicates that a nuclear fusion-based propulsion system is unable to meet these objectives then alternative propulsion concepts may need to be considered. Having a definitive answer to this question, one way or the other, would be a major contribution of the Icarus project to the wider field of interstellar travel studies.

**ACKNOWLEDGEMENTS**

This paper is a submission of the Project Icarus Study Group. It is a revised and updated version of a study initially conducted by the author in April 2012 as part of the Project Icarus Phase IV deliverables for module M14 (Scientific Objectives). I would like to thank Andreas Tziolas for helpful comments on that original report, and Kelvin Long and Rob Swinney for their continued support and leadership of Project Icarus, and interstellar exploration studies more generally.

**REFERENCES**

[1] Long, K.F., Fogg, M.J., Obousy, R., Tziolas, A., Mann, A., Osborne, R. and Presby, A., "Project Icarus: Son of Daedalus – Flying Closer to Another Star", *JBIS*, 62, 403-414, (2009).

[2] Long, K.F., Obousy, R. and Tziolas, A., "Project Icarus: The Origins and Aims of the Study", JBIS, 64, 88-91, (2011).

[3] Crawford, I.A., "The Astronomical, Astrobiological and Planetary Science Case for Interstellar Spaceflight", JBIS, 62, 415-421, (2009).

[4] Webb, G.M., "Project Daedalus: Some Principles for the Design of a Payload for a Stellar Flyby Mission", Project Daedalus: Final Report, *JBIS*, S149-S161, (1978).

[5] Crawford, I.A., "Project Icarus: A Review of Local Interstellar Medium Properties of Relevance for Space Missions to the Nearest Stars", *Acta Astronautica*, 68, 691-699, (2011).

[6] ESA Cassini-Huygens webpage http://www.esa.int/esaMI/Cassini-Huygens/SEMY182VQUD_0.html (accessed 16 January 2016). See also https://en.wikipedia.org/wiki/Cassini%E2%80%93Huygens (accessed 16 January 2016).

[7] McNutt, R.L., Andrews, G.B., McAdams, J., Gold, R.E., Santo, A., Oursler, D., Heeres, K., Fraeman, M. and Williams, B., "Low Cost Interstellar Probe", *Acta Astronautica,* **52**, 267-279, (2003).

[8] The Alpha Magnetic Spectrometer Experiment webpage http://www.ams02.org/ (accessed 16 January 2016).

[9] Fleck, B., Müller, D., Haugan, S., Sánchez Duarte, L., Siili, T. and Gurman, J. B., "10 Years of SOHO", *ESA Bull*., **126**, 24-32, (2006). See also http://en.wikipedia.org/wiki/Solar_and_Heliospheric_Observatory (accessed 16 January 2016).

[10] Harrison, R.A., Davies, J.A., Rouillard, A.P., Davis, C.J., Eyles, C.J.;, Bewsher, D., Crothers, S.R., Howard, R.A., Sheeley, N.R., Vourlidas, A., Webb, D.F., Brown, D.S. and Dorrian, G.D., "Two Years of the STEREO Heliospheric Imagers: Invited Review", *Solar Phys.*, **256**, 219-237, (2009).

[11] ESA Solar Orbiter webpage http://www.esa.int/Our_Activities/Space_Science/Solar_Orbiter (accessed 16 January 2016).

[12] The Galileo atmospheric entry probe https://en.wikipedia.org/wiki/Galileo_Probe (accessed 16 January 2016).

[13] Crawford, I.A. and Smith, A., "MoonLITE: A UK-Led Mission to the Moon", *Astronomy and Geophysics*, **49**, 3.11-3.14, (2008).

[14] Smith, A., et al., "LunarEX: A Proposal to Cosmic Vision," *Experimental Astronomy*, **23**, 711-740, (2009).

[15] Gowen, R. et al., "Penetrators for *In Situ* Sub-Surface Investigations of Europa," *Adv. Space Res.*, **48**, 725-742, (2011).

[16] Grotzinger, J.P., et al., "Mars Science Laboratory Mission and Science Investigation", *Space Science Rev.*, **170**, 5-56, (2012). See also https://en.wikipedia.org/wiki/Mars_Science_Laboratory (accessed 16 January 2016).

[17] Farley, K.A., et al., "*In Situ* Radiometric and Exposure Age Dating of the Martian Surface", Science, **343**, 1247166, (2014).

[18] Rayman, M.D., Fraschetti, T.C., Raymond, C.A. and Russell, C.T., "Dawn: A Mission in Development for Exploration of Main Belt Asteroids Vesta and Ceres", *Acta Astronautica,* **58**, 605- 616, (2006). See also https://en.wikipedia.org/wiki/Dawn_(spacecraft) (accessed 16 January 2016).

[19] ESA Rosetta webpage http://www.esa.int/Our_Activities/Space_Science/Rosetta (accessed 16 January 2016). For Philae, see also https://en.wikipedia.org/wiki/Philae_(spacecraft) (accessed 16 January 2016).

[20] Brown, F.S., Adelson, H.E., Chapman, M. C., Clausen, O.W., Cole, A.J., Cragin, J.T., Day, R.J., Debenham, C.H., Fortney, R.E. and Gilje, R.I, "The Biology Instrument for the Viking Mars Mission", *Rev. Sci. Instruments*, **49**, 139-182, (1978).

[21] Staufer, U., Parrat, D., Gautsch, S., Pike, W.T., Marshall, J., Blaney, D., Mogensen, C. T. and Hecht, M, "The PHOENIX Microscopy Experiments", Fourth International Conference on Mars Polar Science and Exploration, Davos, Switzerland, abstract No. 8097, (2006).

[22] Mahaffy, P.R., et al., "The Sample Analysis at Mars Investigation and Instrument Suite", *Space Science Rev.*, **170**, 401-478, (2012). See also NASA SAM webpage http://msl-scicorner.jpl.nasa.gov/Instruments/SAM/ (accessed 16 January 2016).

[23] Aubrey, A.D., et al., "The Urey Instrument: An Advanced *In Situ* Organic and Oxidant Detector for Mars Exploration", *Astrobiology*, **8**, 583-595, (2008).

[24] Angelo, J.A., *Spacecraft for Astronomy*, Facts on File, New York, (2007).

[25] Baxter, S., "Project Icarus: Exploring Alpha Centauri: Trajectories and Strategies for Subprobe Deployment", JBIS, 69, 11-19, (2016).

[26] Crawford, I.A., "Project Icarus: Astronomical Considerations Relating to the Choice of Target Star", *JBIS*, **63**, 419-425, (2010).

[27] Long, K., Fogg, M., Obousy, R., Tziolas, A., Presby, A., Mann, A., Osborne, R., "Project Icarus: Terms of Reference" (2009). See http://www.icarusinterstellar.org/ToRicarus.pdf (accessed 16 January 2016).

**Table 1**: Preliminary list of Icarus science objectives.

| Scientific Area | Area of Investigation | Examples |
|---|---|---|
| 1 Science *en-route* | Outer solar system studies | Heliosphere, Kuiper belt and Oort Cloud |
| | Local interstellar medium | Structure, density, temperature, composition, magnetic fields, etc |
| | Astronomical studies | Parallax measurements; other? |
| | Fundamental physics | Gravitational waves, gravity, dark matter, etc |
| 2 Stellar astrophysics | Outer environment of target star | Astrosphere, dust disk, comets, etc |
| | Target star astrophysics I | Mass, composition, temperature, mag. fields |
| | Target star astrophysics II | Photospheric and coronal activity; long-term monitoring |
| | Target star astrophysics III | Stellar wind/corona composition |
| 3 Planetary science | Terrestrial planets I | Mass, density, magnetic fields |
| | Terrestrial planets II | Atmospheric composition and structure |
| | Terrestrial planets III | Surface geology |
| | Terrestrial planets IV | Internal structure (geophysics) |
| | Giant planets I | Mass, density, magnetic fields |
| | Giant planets II | Atmospheric composition and structure |
| | Asteroids/small bodies I | Numbers, mass, density, composition |
| | Asteroids/small bodies II | In situ dating of primitive meteorites |
| 4 Astrobiology | Astrobiology I | Identification of habitable environments in target system |
| | Astrobiology II | Search for biomarkers |
| | Astrobiology III | Detection of planetary surfaces |
| | Astrobiology IV | Life detection below planetary surfaces |
| | Astrobiology V | Search for evidence of extinct life |
| | Astrobiology VI | Biochemical characterization of extant life forms |
| | Astrobiology VII | Search for evidence of technological artefacts in target system |

**Table 2** : Summary of types, and estimated numbers, of sub-probes required to conduct an exploration of a multi-planet system sufficient to meet the Project Icarus science objectives identified in Table 1. An additional 10 tonnes are estimated for cruise-phase instruments (science area #1). See text for discussion.

| Scientific Area | Probe Type | Example | Probe dry mass (tonnes) | Number of probes required | Total probe dry mass (tonnes) |
|---|---|---|---|---|---|
| 2 | Stellar orbiters | SOHO | 1.35 | 2 - 4 | 2.7 – 5.4 |
| 3 | Planetary orbiters | Cassini | 2.15 | 6 - 8 | 13 – 17 |
| 3 | Gas Giant probes/balloons | Galileo entry probe | 0.34 | 2 - 4 | 0.7 – 1.4 |
| 3/4 | Rocky Planet atmospheric probes | Galileo entry probe | 0.34 | 6 - 8 | 2.0 – 2.7 |
| 3/4 | Planetary soft landers | MSL | 0.90* | 6 - 8 | 5.4 – 7.2 |
| 2/3 | Minor body (asteroid/comet) probes | Rosetta | 1.33 | 2 - 4 | 2.7 – 5.3 |
| 3/4 | Low mass penetrators | Lunar-EX | 0.04 | 30 - 50 | 1.2 – 2.0 |
| | **Total** | | | | **28 – 41** |

* Rover mass only, excluding mass required for entry, descent and landing.